# Molecular Dynamics Study of Polarizable Ion Models for Molten AgBr

Vicente Bitrián and Joaquim Trullàs

*Departament de Física i Enginyeria Nuclear, Universitat Politècnica de Catalunya, Campus Nord UPC, 08034 Barcelona, Spain.*

Three different polarizable ion models for molten AgBr have been studied by molecular dynamics simulations. The three models are based on a rigid ion model (RIM) with a pair potential of the type proposed by Vashishta and Rahman for $\alpha$-AgI, to which the induced dipole polarization of the ions is added. In the first (PIM1) the dipole moments are only induced by the local electric field, while in the other two (PIM1s and PIM2s) a short-range overlap induced polarization opposes the electrically induced dipole moments. In the PIM1 and the PIM1s only the anions are assumed polarizable, while in the PIM2s both species are polarizable. Long molecular dynamics simulations show that the PIM2s is an unphysical model since, for some improbable but possible critical configurations, the ions become infinitely polarized. The results of using the PIM1, the PIM1s, as well as those of the simple RIM, have been compared for the static structure and ionic transport properties. The PIM1 reproduces the broad main peak of the total structure factor present in the neutron diffraction data, although the smoothed three-peak feature of this broad peak is slightly overestimated. The structural results for the PIM1s are intermediate between those for the RIM and the PIM1, but fail to reproduce the experimental features within the broad principal peak. Concerning the ionic transport properties, the value of the conductivity obtained using PIM1 is in good agreement with experimental values, while the self-diffusion coefficients and the conductivity for the PIM1s are lower than the corresponding values using the PIM1 and the RIM.

## 1. Introduction

A few years ago Tasseven et al.[1] reported the results of calculations for molten AgBr and AgCl using the hypernetted chain theory of liquids and molecular dynamics (MD) simulations. In these calculations they used a rigid ion model (RIM) with pair potentials of the functional form originally proposed by Vashishta and Rahman for the superionic $\alpha$-AgI and $\alpha$-CuI.[2,3] The results for the total static structure factor $S_T(k)$ were in qualitative agreement with the available experimental neutron scattering data.[4,5] However, the results failed to reproduce the characteristic three-peak structure present in the broad main peak of the experimental $S_T(k)$. In addition, the results for the ionic transport properties resembled the superionic-like behavior found in MD simulations of copper halides and silver iodide melts,[6] with the self-diffusion coefficients for the cations almost double in value than those for the anions, as if the transition to a superionic phase is finally realized in AgCl and AgBr. We recall that Andreoni and Tosi,[7] and later Nield et al.,[8] suggested that melting frustrates a possible transition of AgCl and AgBr to a superionic phase.

At the same time, Wilson et al.[9] showed by MD simulations that the three-peak structure of molten AgCl can be predicted if polarizable ion models are considered. They studied models where the dipole and quadrupole polarization effects are added to the rigid ion pair potential of the Born–Mayer form proposed by Mayer[10] for AgCl. The simple rigid ion potential of Mayer fails to predict the melting point, which is at higher temperature than experimental, but including polarization effects reduces the melting temperature. However, the results for the self-diffusion coefficients were too small and the values of the ionic conductivities estimated from the Nernst–Einstein relation were significantly lower than the experimental data. We note, however, that the purpose of the paper of Wilson et al. was to examine the way in which induced quadrupoles might influence the properties of solid and liquid AgCl rather than to develop an accurate model.

More recently, Trullàs et al.[11,12] showed that the three-peak structure of molten AgCl can also be reproduced if the anion induced dipole polarization contributions are added to the Vashishta–Rahman potentials proposed in ref 1. Furthermore, it was found that the results for the conductivity were in good agreement with experimental values.

The aim of this work is to extend to AgBr the above MD studies of polarizable ions models used for molten AgCl. Although we are interested to see if polarizable effects improve the results for $S_T(k)$, we want to stress that the main purpose is to study the influence of the induced dipole polarization on the properties of molten AgBr. We started with a rigid ion model with the pair potential of the Born–Mayer form proposed by Mayer[10] for AgBr. This potential, as it was the case for molten AgCl, fails to predict the melting point. However, we found that the inclusion of polarization effects does not reduce enough the melting temperature and the system does not melt at the thermodynamic state at which experimental neutron scattering data are available. For this reason the polarizable ion models of this work are restricted to those based on the Vashishta–Rahman potential. In the first case (PIM1), the dipole moments are only induced by the local electric field, while in the other two (PIM1s and PIM2s), a "mechanical" short-range polarizability opposes the electrically induced dipole moments. This second type of polarizability was not studied in the previous work of molten AgCl.[11,12] In the

PIM1 and the PIM1s only the anions are assumed to be polarizable, while in the PIM2s both cations and anions are polarizable. Although preliminary PIM2s simulations with 216 ions were stable over $300\times10^3$ time steps, we found that for longer runs, or simulations with 1000 ions, there are some critical configurations where the ions polarize catastrophically. Thus, in this work we only present the results obtained by using the RIM, the PIM1 and the PIM1s.

The layout of the paper is as follows. We describe the models in section 2. In section 3 we describe the computational details and analyze the polarization catastrophe observed in the PIM2s simulations. In section 4 we present and discuss the results of our simulations. Finally we sum up our results in the concluding remarks of section 5.

## 2. Interaction Models

**2.1. The Rigid Ion Model.** The potential energy of the rigid ion model (RIM) studied in this work for molten AgBr is

$$U^{RIM} = \frac{1}{2}\sum_{i=1}^{N}\sum_{j\neq i}^{N} f_{ij}^{VR}(r_{ij}) \quad (2.1)$$

where the functional form of the effective pair potential is that originally proposed by Vashishta and Rahman,[2]

$$f_{ab}^{VR}(r) = f_{ab}^{0}(r) - \frac{P_{ab}}{r^4} \quad (2.2)$$

with

$$f_{ab}^{0}(r) = \frac{z_a z_b e^2}{r} + \frac{H_{ab}}{r^n} - \frac{C_{ab}}{r^6} \quad (2.3)$$

where we use the $a$ and $b$ subscripts to denote species rather than particles. The first term on the right hand side of eq (2.3) is the Coulomb interaction between ionic charges, with $z_a<1$ the effective charge in units of the fundamental charge $e$; the second models the repulsion between the ions arising from the overlap of the outer shell of electrons, with $H_{ab} = A(s_a+s_b)^n$, where $s_a$ are related to the ionic radii, $A$ describes the repulsive strength and $n$ the hardness; and the third is the van der Waals contribution, with $C_{ab}=(3/2)\alpha_a\alpha_b(E_a^{-1}+E_b^{-1})^{-1}$, where $\alpha_a$ are the electronic polarizabilities and $E_a$ are related to the ionization potentials of the cations and electron affinities of the anions. The last term in eq (2.2) is the effective monopole-induced dipole attractive interaction, with $P_{ab}=(1/2)(\alpha_a z_b^2 + \alpha_b z_a^2)e^2$.

In this work we use $n = 7$ for the AgBr repulsive potential. This value is the same as that used in the original formulation of Vashishta and Rahman[2] and in a recent study of the behavior of solid AgI and CuI under pressure.[13] This value differs from the $n = 6$ used by Tasseven et al.[1] However, in a recent study[14] of the structure of the molten mixture Ag($I_{0.3}Br_{0.7}$) we re-examined the parameterization of the AgBr potential using $n = 7$ and we noted that it does not affect in any significant way the results for the structure of molten AgBr. We parameterized the AgBr interaction potential following the prescription suggested by Rahman and Vashishta for AgI.[15] We assumed $|z_a| = 0.66$ (from ref 16), $\alpha_{Ag} = 0$ (thus $P_{AgAg} = 0$ and $C_{AgAg} = C_{AgBr} = 0$), $\alpha_{Br} = 4.16$ Å$^3$ (from ref 17) and $C_{BrBr} = 112.6$ eVÅ$^6$ (with $E_{Br} = 8.67$ eV from ref 10), the same values proposed by Tasseven et al.[1] for AgBr. The values of $s_{Ag}$ and $s_{Br}$, which appear as scale factors in $H_{ab}$, were determined assuming that $s_{Br}+s_{Ag}=a_0/2$ and $2s_{Br}=a_0\sqrt{2}/2$ are the nearest neighbor distances in a rock-salt structure with the lattice constant $a_0 = 5.7745$ Å of the AgBr crystal.[18] We estimated the value of $A$ for $n =7$, and therefore the $H_{ab}$ values, from the condition that the crystal energy minimum occurs at the distance $a_0$. However, since MD simulations showed that the estimated value of $A$ is too high and the system does not melt at the required temperature, we lowered its value until the system melts. The final values of potential parameters in eqs (2.2) and (2.3) for molten AgBr that we have used in this work are shown in Table 1.

**TABLE 1: Potential Parameters of the RIM, eq (2.2), for Molten AgBr, with $|z_a| = 0.66$ and n = 7**[a]

|  | Ag$^+$–Ag$^+$ | Ag$^+$–Br$^-$ | Br$^-$–Br$^-$ |
|---|---|---|---|
| $H_{ab}$ / eVÅ$^7$ | 5.11 | 216 | 2440 |
| $P_{ab}$ / eVÅ$^4$ | 0 | 13.046 | 26.092 |
| $C_{ab}$ / eVÅ$^6$ | 0 | 0 | 112.6 |

[a] In the polarizable ion models $P_{ab} = 0$ for all interactions.

**2.2. The Polarizable Ion Models.** The polarizable ion models are constructed by adding the induced polarization contributions to the pair potential $f_{ab}^0(r)$. We assume that two types of dipoles are induced in an ion placed at position $\mathbf{r}_i$. The first type is the dipole induced by the local electric field $\mathbf{E}_i$ due to all other ions, whose moment $\mathbf{p}_i$ is given by the polarizability $\alpha_i$ in the linear approximation $\mathbf{p}_i = \alpha_i\mathbf{E}_i$. If only this type of polarization is introduced, the ions may become over-polarized and the polarization catastrophe takes place (see below). The second type is the deformation dipole induced by short-range overlap effects due to the neighboring ions. For the resulting dipole moment we use the following constitutive relation

$$\mathbf{p}_i = \alpha_i\mathbf{E}_i + \alpha_i\sum_{j\neq i}^{N} s_{ij}(r_{ij})\mathbf{r}_{ij} \quad (2.4)$$

where the short-range effects in the second term are approximated to be additive. The local field at $\mathbf{r}_i$ is

$$\mathbf{E}_i = \mathbf{E}_i^q + \mathbf{E}_i^p = \sum_{j\neq i}^{N}\frac{z_j e}{r_{ij}^3}\mathbf{r}_{ij} + \sum_{j\neq i}^{N}\left(3\frac{(\mathbf{p}_j\cdot\mathbf{r}_{ij})}{r_{ij}^5}\mathbf{r}_{ij} - \frac{1}{r_{ij}^3}\mathbf{p}_j\right) \quad (2.5)$$

where $\mathbf{r}_{ij} = \mathbf{r}_i - \mathbf{r}_j$, $\mathbf{E}_i^q$ denotes the field at $\mathbf{r}_i$ due to all the point charges except $q_i = z_ie$, and $\mathbf{E}_i^p$ denotes the field at $\mathbf{r}_i$ due to all the dipole moments except $\mathbf{p}_i$. Substitution of eq (2.5) into eq (2.4) leads to a system of linear equations that determines uniquely the $N$ dipole moments $\{\mathbf{p}_i\}$ for given ionic positions $\{\mathbf{r}_i\}$.

Following Wilson and Madden[19] we use

$$s_{ab}(r) = -f_{ab}(r)\frac{z_b e}{r^3} \quad (2.6)$$

where $f_{ab}(r)$ is a suitable short-range function only effective over length scales corresponding to the nearest-neighbor separation and limiting value 1 as $r\to 0$. By choosing this form we ensure that the short-range terms cancel the $(z_j e/r_{ij}^3)\mathbf{r}_{ij}$ contributions of $\mathbf{E}_i^q$ in eq (2.4), if two ions approach unphysical separations. A convenient form for $f_{ab}(r)$ is the Tang and Toennies dispersion damping function,[20]

$$f_{ab}(r) = \exp(-r/\boldsymbol{r}_{ab}) \sum_{k=0}^{4} \frac{(r/\boldsymbol{r}_{ab})^k}{k!} \quad (2.7)$$

where the single parameter $\boldsymbol{r}_{ab}$ is the length scale over which the overlap damping acts. The short-range polarization effects are eliminated in the limit $\boldsymbol{r}_{ab} = 0$.

The potential energy of the above polarizable ion model may be conveniently written as

$$U^{\text{PIM}} = \frac{1}{2} \sum_{i=1}^{N} \sum_{j \neq i}^{N} f_{ij}^0(r_{ij}) - \sum_{i=1}^{N} \mathbf{p}_i \cdot \mathbf{E}_i^q - \frac{1}{2} \sum_{i=1}^{N} \mathbf{p}_i \cdot \mathbf{E}_i^p + \sum_{i=1}^{N} \frac{p_i^2}{2\boldsymbol{a}_i} - \sum_{i=1}^{N} \sum_{j \neq i}^{N} s_{ij}(r_{ij}) \mathbf{p}_i \cdot \mathbf{r}_{ij} \quad (2.8)$$

The first term corresponds to the potential energy of a rigid ion model which ions interact via the pair potential $f_{ij}^0$ of eq (2.3); the second term is the energy contribution of the charge-dipole interactions; the third is the energy of the dipole-dipole interaction; the fourth is the self-energy contribution, that is, the internal energy of the $N$ induced dipoles, which corresponds to the energy required in creating them; and the last is the contribution due to the short-range damping interactions.[9] The second, third and fourth terms give the usual potential energy of the dipoles linearly induced by the electric field.[21,22,23] However, when short-range polarization effects are included, the last term must be added to $U^{\text{PIM}}$. This term is derived taking into account the constitutive relation (2.4) and that the induced dipoles minimize $U^{\text{PIM}}$, i.e. the gradient of $U^{\text{PIM}}$ with respect to each dipole moment must vanish.

In carrying out the MD simulations we actually have to calculate the forces. This is done taking into account that the dipole moments adjust themselves to minimize $U^{\text{PIM}}$, so that the gradient of $U^{\text{PIM}}$ with respect to $\mathbf{p}_j$ vanishes. Although the dipole moments are determined by the field, and therefore depend on the ionic positions, the gradient calculations are simplified by the minimization condition, which implies that only explicit derivatives of $U^{\text{PIM}}$ with respect to the coordinates have to be considered. Then, the force acting on the ion $i$ can be written as in ref 12 plus an extra term due to the short-range damping interactions given by

$$\mathbf{F}_i^{ps} = \sum_{j \neq i}^{N} \left( \frac{\dot{s}_{ij}(r_{ij})}{r_{ij}} (\mathbf{p}_i \cdot \mathbf{r}_{ij}) \mathbf{r}_{ij} - \frac{\dot{s}_{ji}(r_{ij})}{r_{ij}} (\mathbf{p}_j \cdot \mathbf{r}_{ij}) \mathbf{r}_{ij} + s_{ij}(r_{ij}) \mathbf{p}_i - s_{ji}(r_{ij}) \mathbf{p}_j \right) \quad (2.9)$$

where $\dot{s}_{ij}(r) = ds_{ij}(r)/dr$.

We have simulated three polarizable ion models in this work. In the first, which we denote PIM1, the dipole moments are only induced by the local electric field, i.e. $\boldsymbol{r}_{ab} = 0$ and, thus, $s_{ab}(r) = 0$. In the other two, which we denote PIM1s and PIM2s, dipoles induced by short-range overlap effects oppose the electrically induced dipole moments with $\boldsymbol{r}_{ab} = 0.5$ Å, a slightly higher value than that proposed by Wilson et al.[9] for AgCl because $s_{\text{Br}} > s_{\text{Cl}}$. In the PIM1 and the PIM1s only the anions are assumed polarizable with $\boldsymbol{a}_- = 4.16$ Å$^3$ and $\boldsymbol{a}_+ = 0$, the same values used for the RIM parameterization; while in the PIM2s both species are polarizable with $\boldsymbol{a}_+ = 1.67$ Å$^3$ as in ref 16. Summing up, the three polarizable ion models are characterized by the pair potentials $f_{ij}^0$ of eq (2.3) with the same parameter values given in Table I for the RIM, and the polarizabilities and the short-range damping parameters given in Table 2.

**TABLE 2: Polarizabilities and Short-Range Damping Parameters of the Polarizable Ion Models**

|  | PIM1 | PIM1s | PIM2s[a] | PIM2[a] |
|---|---|---|---|---|
| $\alpha_+ = \alpha_{\text{Ag}}$ | 0 | 0 | 1.67 Å$^3$ | 1.67 Å$^3$ |
| $\alpha_- = \alpha_{\text{Br}}$ | 4.16 Å$^3$ | 4.16 Å$^3$ | 4.16 Å$^3$ | 4.16 Å$^3$ |
| $\boldsymbol{r}_{ab}$ | 0 | 0.5 Å | 0.5 Å | 0 |

[a] PIM2 and PIM2s are unphysical models since lead to the polarization catastrophe.

We will also refer to PIM2 as the model where both species are electrically polarizable without short-range effects. The PIM2 is an unphysical model since, at short enough distances, two unlike ions polarize catastrophically –namely, the ions become infinitely polarized– as it will be shown below. Furthermore, MD simulations carried out for this work have shown that the PIM2s is unstable. As we discuss in the next section, for some improbable but possible critical configurations the ions become over-polarized and the polarization catastrophe occurs.

**2.3. The Potential Energy of Two Isolated Ions and The Polarization Catastrophe.** For two isolated ions polarized by the local electric field and the short-range damping interactions, the system of eqs (2.4) and (2.5) can be solved analytically. Then, it is found that the dipole moment modulus induced in the $a$-type ion by the $b$-type ion at distance $r$ is

$$p_a(r) = \left| \frac{\boldsymbol{a}_a}{1-(r_c/r)^6} [1-f_{ab}(r)] \left( \frac{z_b e}{r^2} - \frac{2\boldsymbol{a}_b z_a e}{r^5} \right) \right| \quad (2.10)$$

and the corresponding potential energy, eq (2.8), for two ions with opposite charge can be written as

$$f_{+-}^{\text{PIM}}(r) = f_{+-}^0(r) - [1-f_{+-}(r)]^2 \frac{\frac{1}{2}[(\boldsymbol{a}_+ z_-^2 + \boldsymbol{a}_- z_+^2) r^3 - z_+ z_- r_c^6] e^2}{r(r^3 + r_c^3)(r^3 - r_c^3)} \quad (2.11)$$

where $r_c = (4\boldsymbol{a}_+ \boldsymbol{a}_-)^{1/6}$ is a singular point we will call the *polarization catastrophe distance*. If two ions with opposite charge approach each other at the distance $r_c$, they become infinitely polarized and the induced attractive interaction is infinite, namely the polarization catastrophe occurs. However, there is no polarization catastrophe distance between like ions. When $a = b$, eq (2.10) can be written as

$$p_a(r) = \left| \boldsymbol{a}_a [1-f_{aa}(r)] \frac{z_a e}{r^3 + 2\boldsymbol{a}_a} r \right| \quad (2.12)$$

without a singular point. For this reason we have written eq (2.11) with the subscripts $+$ and $-$ instead of two dummy subscripts as $a$ and $b$. If the subscripts $+$ and $-$ in eq (2.11) are replaced by $a$ and $b$, it is easy to see that for like ions ($a = b$) the factor $(r^3 - r_c^3)$ in the denominator also appears in the numerator and cancels itself, and the potential energy for like ions can be written as

$$f_{aa}^{PIM}(r) = f_{aa}^0(r) - [1 - f_{aa}(r)]^2 \frac{a_a z_a^2 e^2}{r(r^3 + 2a_a)} \quad (2.13)$$

without a polarization catastrophe distance.

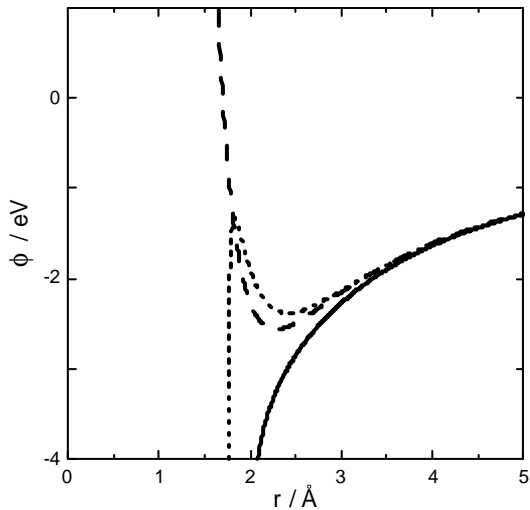

**Figure 1.** Potential energy of two isolated ions Ag$^+$ and Br$^-$: $f_{+-}^{PIM}(r)$ given by eq. (2.11) with $a_+ = 1.67$ Å$^3$, $a_- = 4.16$ Å$^3$, and $r_{+-} = 0$ ($f_{+-}(r) = 0$), as in the PIM2 (solid line); $f_{+-}^{PIM}(r)$ given by eq. (2.11) with $a_+ = 1.67$ Å$^3$, $a_- = 4.16$ Å$^3$, and $r_{+-} = 0.5$ Å ($f_{+-}(r) \neq 0$), as in the PIM2s (dotted line); $f_{+-}^{PIM}(r)$ for $a_+ = 0$ given by eq (2.14) with $a_- = 4.16$ Å$^3$ and $r_{+-} = 0$ ($f_{+-}(r) = 0$), as in the PIM1 (dashed line).

In Figure 1 we plot $f_{+-}^{PIM}(r)$ for two isolated ions with $a_+ = a_{Ag} = 1.67$ Å$^3$ and $a_- = a_{Br} = 4.16$ Å$^3$, so that $r_c = 1.74$ Å, and the potential parameters for $f_{+-}^0(r)$ given in Table 1. When the short-range polarization effects are neglected (solid line), that is $r_{+-} = 0$ and thus $f_{+-}(r) = 0$, $f_{+-}^{PIM}(r)$ is more attractive as the ions approach at the polarization catastrophe distance $r_c$, where it goes to $-\infty$. The shape of $f_{+-}^{PIM}(r)$ for this case clearly illustrates why the PIM2 is an unphysical model. However, when the short-range damping interactions are included, as in the PIM2s, the polarization contributions of the second term in eq (2.11) are less intense at short distances and the repulsion term of $f_{+-}^0(r)$ can avoid that two unlike ions approach at $r_c$. As seen in Figure 1, for $r_{+-} = 0.5$ Å (dotted line) $f_{+-}^{PIM}(r)$ presents a repulsive barrier from 1.83 Å to 2.41 Å. For $r_{+-}$ values lower than 0.5 Å the repulsive barrier becomes less intense and can disappear. We note that for $r_c$ small enough, much lower than the ionic diameters, the repulsion term of $f_{+-}^0(r)$ could avoid the polarization catastrophe without the short-range damping interactions.

Despite the above discussion about two isolated ions with opposite charge, MD simulations carried out in this work for the PIM2s in condensed phase have shown that for some critical configurations the ions can become over-polarized due to many-body effects. This can be understood by arguing that the many-body effects reduce the repulsive barrier of $f_{+-}^{PIM}(r)$. If we assume two isolated polarizable ions with a permanent dipole moment, which in some sense represents the environmental effects, the potential energy will be given by eq (2.11) plus extra terms corresponding to the permanent dipoles contributions that change the shape of $f_{+-}^{PIM}(r)$.

We now turn to analyze the potential energy of two isolated ions with opposite charge when only the anion is polarizable, as in the case of the PIM1 and the PIM1s. Here, there is no a polarization catastrophe distance between the two ions. If $a_+ = 0$, $r_c = 0$, and eq (2.11) reads

$$f_{+-}^{PIM}(r) = f_{+-}^0(r) - [1 - f_{+-}(r)]^2 \frac{\frac{1}{2} a_- z_+^2 e^2}{r^4}$$

$$(\text{if } a_+ = 0) \quad (2.14)$$

Hence, if the short-range damping interactions are neglected, that is $f_{+-}(r) = 0$, as in the case of the PIM1, eq (2.14) becomes $f_{+-}^{VR}(r)$. In Figure 1, we also plot $f_{+-}^{VR}(r)$. We note that, when $f_{+-}(r) \neq 0$, the repulsive wall and the minimum of the potential given by eq (2.14) are shifted to larger distances compared to those for $f_{+-}^{VR}(r)$. This is the case of the PIM1s.

### 3. Computational Details

We have studied molten AgBr by using the simple rigid ion model (RIM) and the three polarizable ion models (PIM1, PIM1s and PIM2s) described in the above section. MD simulations have been carried out considering the ions placed in a cubic box with periodic boundary conditions, and using the Beeman's algorithm with a time step of $\Delta t = 5 \times 10^{-15}$ s. The electric fields $\mathbf{E}_i^q$ and $\mathbf{E}_i^p$, and the corresponding long range interactions between charges and induced dipole moments —first term in eq (2.3) and second and third terms in eq (2.8), as well as the corresponding forces— have been evaluated by the Ewald method. Details of the Ewald sums used in our simulations are given in ref 11. The Ewald parameter has been chosen to be $5/L$ (where $L$ is the length of the side box), the real space terms have been truncated at distances longer than $L/2$, and around 180 wave-vectors have been used in the reciprocal space contributions.

The system of eqs (2.4) and (2.5), whose solution gives the dipole moments for given ionic positions, may be solved by using the matrix inversion method.[21,24] However, since this method is too expensive computationally, the dipole moments have been evaluated by using the prediction-correction iterative method proposed by Vesely.[25,12] In this iterative method, an initial guess for the fields $\{\mathbf{E}_i^p\}$ is made by using the dipole moments from the previous time step, and the dipole moments resulting from these fields are evaluated using eq (2.4), which can be iterated to self-consistency. We have used a convergence limit of $|\Delta\mathbf{E}_i^p|^2/|\mathbf{E}_i^p|^2 < 10^{-4}$ for each ion, instead of their sum $\Sigma(|\Delta\mathbf{E}_i^p|^2/|\mathbf{E}_i^p|^2)$ as we used in ref 11. Then, in the PIM1 and PIM1s simulations, the convergence is reached after about 7 or 8 iterations, while in the PIM2s simulations more than 40 iterations can be required.

For the systems under study the initial conditions in the equilibration process are important, and it is essential to start with an initial configuration in which the anions exhibit liquid behavior. We chose a distorted rock-salt crystalline structure at an ionic number density $r_N$ smaller and at a temperature $T$ higher than the density and temperature of interest, so that ions diffuse more rapidly, thus ensuring that equilibrium was achieved in a relatively short time. Once equilibrium was reached, we compressed and cooled the system in several steps, making sure that at each step the liquid behavior was preserved, until reaching the density and temperature of interest. It is worth mentioning that the equilibration process required long simulations to make sure that the system did not quench into a solid phase, as it did when we studied polarizable ions models based on the pair potential of the Born–Mayer.[10]

As in ref 1, we have simulated molten AgBr at 883 K and 753 K, the temperatures at which Inui et al.[5] reported the neutron scattering data, with ionic number densities 0.0346 ions/Å$^3$ and 0.0354 ions/Å$^3$, respectively.[26] At both temperatures we have carried out two sets of simulations averaging over $300 \times 10^3$ configurations after equilibrium was achieved: a preliminary set using $N$ = 216 ions ($N_+$ = 108 cations and $N_-$ = 108 anions); and a second set with $N$ = 1000 in order to reduce the statistical fluctuations in the static structure factors at the wave numbers of the broad main peak of $S_T(k)$. The radial distribution functions and the time correlations functions obtained with 216 and 1000 ions are practically identical to the eye, while the discrepancies in the values of the transport coefficients are around 5%. Furthermore, the differences with temperature are those expected, i.e. at the higher temperature the peaks of the radial distribution functions and the static structure factors are lower and slightly broader, while the self-diffusion coefficients and the conductivities are larger. Since there are also experimental $S_T(k)$ measured at 703 K by Keen et al.[27] near the melting point (701 K), and at 773 K by Takeda et al.,[28] in this paper we only present the results corresponding to molten AgBr at 753 K. Moreover, since the PIM2s leads to the polarization catastrophe, in section 4 we only present the results for the RIM, the PIM1 and the PIM1s.

**3.1. The PIM2s and the Polarization Catastrophe.** MD simulations using the PIM2s require special attention. Despite that our preliminary PIM2s simulations at 883 K with 216 ions were stable over $300 \times 10^3$ time steps (1.5 ns), we found that simulations with 1000 ions were stable only over $20 \times 10^3$ time steps. Later, we found that simulations with 216 ions also failed after much longer runs.

To understand the PIM2s instabilities, we carried out a careful check of the configurations at each time step. We saw that, for some improbable critical configurations, two unlike ions are too polarized, and the iterative method to determine the induced dipole moments does not converge, as it leads to unstable oscillations in the predicted dipoles. To crosscheck this method, starting from a non-critical configuration, we carried out MD simulations solving the system of eqs (2.4) and (2.5) by using the exact method of matrix inversion.[21,24] For non-critical configurations we verified the self-consistency of the iterative method, since both matrix inversion and iterative methods gave the same results. Once the critical configuration was reached, the results from the matrix inversion method showed that the polarization of all ions increased faster from one step to the next until the dipole moments were practically infinite.

We interpret the polarization catastrophe as follows. Since the short-range damping interactions used in this work only oppose the electric field due the neighboring monopoles, $(z_j e / r_{ij}^3) \mathbf{r}_{ij}$, they do not cancel the electric field $\mathbf{E}_i^p$ due to the induced dipoles. Thus, for some critical configurations, two neighboring unlike ions can have too large dipole moments that, in turn, induce large dipole moments to the other neighboring ions and, in a feedback process, all ions become over-polarized and the polarization catastrophe take place.

It is worth mentioning that the above PIM2s over-polarization process is different from that observed for unstable MD simulations of the PIM2. Here, for any initial configuration, the dipole moments of the two nearest unlike ions go to infinity as their separation approaches the critical distance, independently of the neighboring ions. While the former is a many-body process, the latter is practically the two-body process predicted by eq (2.11). Furthermore, MD simulations showed that the iterative method converges during the PIM2 over-polarization process.

MD simulations of molten AgCl carried out by Wilson et al.[9] using a PIM2s (PIM2 in their notation) did not detect the polarization catastrophe. This may be due to either the following three reasons: (a) they only used 216 ions and their runs were not long enough to detect a critical configuration; (b) they use an extended Lagrangian approach –in which each dipole is treated as a dynamical variable–[21] that could avoid critical configurations; and (c) there are no critical configurations in the PIM2s they used for molten AgCl. It may be worth investigating further these conjectures.

### 4. Results

**4.1. Liquid Structure.** The basic structural properties calculated in the simulations are the partial radial distribution functions $g_{ab}(r)$ and the Ashcroft–Langreth[29,30] partial structure factors $S_{ab}(k)$, from which the total static structure factor $S_T(k)$ has been evaluated, as well as the Bhatia–Thornton[31,30] structure factors $S_{NN}(k)$, $S_{ZZ}(k)$ and $S_{NZ}(k)$. We adopt a hybrid method for the calculation of the partial structure factors.[12] For wave-numbers lower than 4 Å$^{-1}$ the calculation have been carried out directly from the simulations using

$$S_{ab}(k) = \frac{1}{n_k} \sum_{|\mathbf{k}|=k} \left\{ \frac{1}{N\sqrt{x_a x_b}} \left\langle \sum_{i_a=1}^{N_a} \exp(i\mathbf{k} \cdot \mathbf{r}_{i_a}) \sum_{j_b=1}^{N_b} \exp(-i\mathbf{k} \cdot \mathbf{r}_{j_b}) \right\rangle \right\} \quad (4.1)$$

where the brackets denote the ensemble average over the equilibrium configurations, $\mathbf{r}_{i_a}$ is the position of the particle $i_a$ of species $a$, $\mathbf{k} = (2\pi/L)\mathbf{n}$ is the wave-vector with $\mathbf{n}$ a vector with integer components, $n_k$ is the number of wave-number vectors with the same modulus $k = |\mathbf{k}|$, and $x_a = N_a/N$ is the ionic fraction ($x_+ = x_- = 0.5$). On the other hand, for larger wave-numbers $S_{ab}(k)$ have been evaluated by the Fourier inversion of $h_{ab}(r) = g_{ab}(r) - 1$,

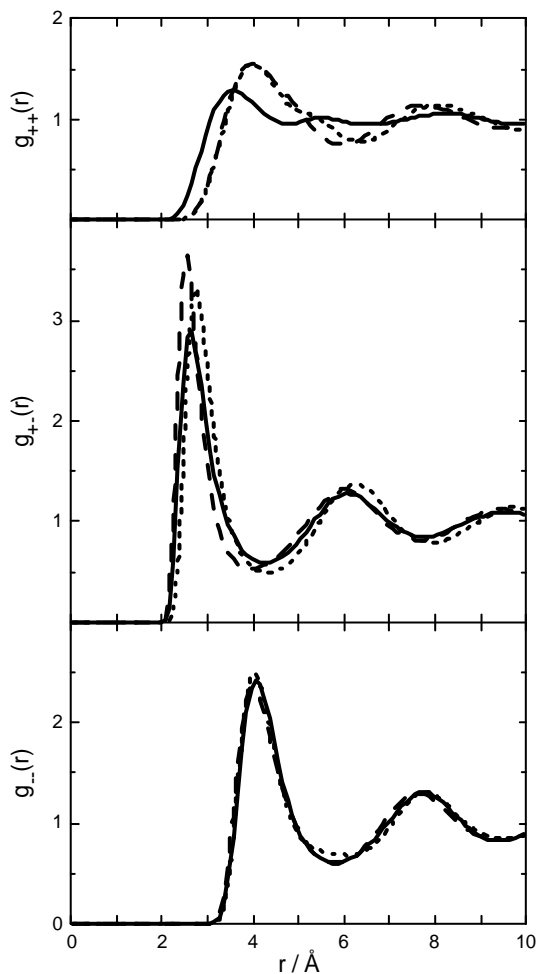

**Figure 2.** Radial distribution functions, $g_{++}(r)$ (top), $g_{+-}(r)$ (middle) and $g_{--}(r)$ (bottom), from MD simulations using the following models: RIM (dashed line), PIM1 (solid line) and PIM1s (dotted line).

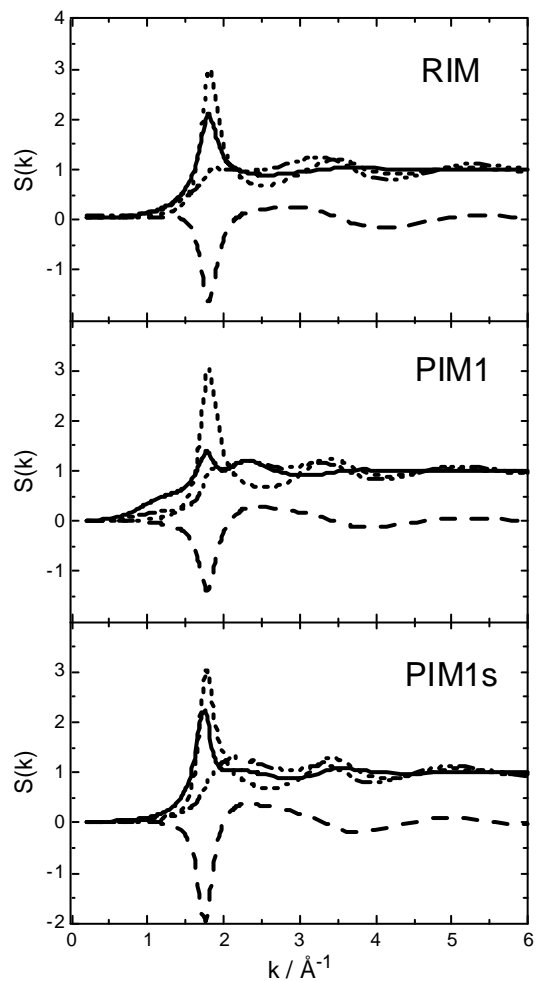

**Figure 3.** Ashcroft–Langreth partial structure factors, $S_{++}(k)$ (solid line), $S_{--}(k)$ (dotted line), $S_{+-}(k)$ (dashed line), and the total structure factor $S_T(k)$ (dash-3-dots line), from MD simulations using the following models: RIM (top), PIM1 (middle) and PIM1s (bottom).

$$S_{ab}(k) = \delta_{ab} + \rho_N \sqrt{x_a x_b} \int_0^\infty [g_{ab}(r) - 1] \frac{\sin(kr)}{kr} 4\pi r^2 dr \quad (4.2)$$

where $\delta_{ab}$ is the Kronecker delta and $\rho_N = N/V$ is the ionic number density (with $V = L^3$).

The radial distribution functions $g_{ab}(r)$ calculated for RIM, PIM1 and PIM1s, as well as for the PIM2s stable configurations averaged before the system becomes unstable, show that the dominant feature of their structure is the charge ordering characteristic of simple molten salts, i.e. the alternation of concentric shells of oppositely charged ions ($g_{++}$ and $g_{--}$ oscillate in antiphase to $g_{+-}$). Moreover, they present the two distinctive features of the structure of molten silver and copper halides,[1,32] namely: (a) the asymmetry between $g_{++}$ and $g_{--}$, with the oscillations of $g_{++}$ less marked than those of $g_{--}$; and (b) the deep penetration of the cations into the first coordination shell, specifically the penetration of the low $r$ tail of $g_{++}$ beneath the principal peak of $g_{+-}$.

In Figure 2 we compare the radial distribution functions $g_{ab}(r)$ calculated at 753 K using the RIM, the PIM1 and the PIM1s. The main differences observed between the $g_{++}$ for PIM1 and those for RIM and PIM1s are: (a) the former exhibits less structure, the peaks are lower and valleys are shallower; (b) it is not in phase with $g_{--}$; (c) its first peak position is shifted to lower values of $r$, from 3.95 Å for RIM and PIM1s to 3.55 Å for PIM1, with a deeper cations penetration; (d) it exhibits a second peak, at 5.5 Å, between the first and second peaks of $g_{+-}$, i.e. around each cation there is a double shell of cations between the first and second shells of anions; and (e) its third peak position, at 8.25 Å, is at larger values of $r$ than the second peak position of $g_{--}$, at 7.75 Å. It is interesting to note that, although we only take into account the anion polarizability, it is the PIM1 cation-cation structure that is mainly affected by the induced polarization interactions, while the anion-anion structure shown by $g_{--}$ is practically the same for the three models, with the first minimum of the $g_{--}$ for PIM1s slightly shallower.

The above features of the PIM1 $g_{++}$ may be attributed to the screening of the cations repulsion due to the anion induced dipoles, namely the negative ends of the anion dipoles attract the cations and, therefore, the separation between the cations can be smaller than it would be the case if the anions were not polarized. The similarity of the $g_{--}$ for the three models may be understood as a packing effect due to the large size of the anions.

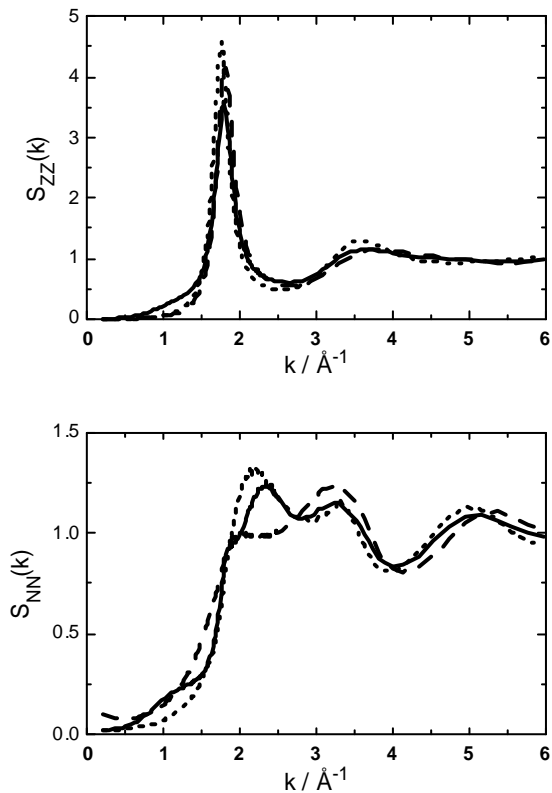

**Figure 4.** Bhatia–Thornton structure factors, $S_{ZZ}(k)$ (top) and $S_{NN}(k)$ (bottom), from MD simulations using the following models: RIM (dashed line), PIM1 (solid line) and PIM1s (dotted line). Note the change in scaling.

The similarity between the structure of the RIM and the PIM1s indicates that the short-range damping interactions included in PIM1s partially cancel the effects of the electrically induced dipoles. However, the following differences are observed: (a) the $g_{++}$ for PIM1s shows a shoulder at about the same position of the PIM1 second peak, and its first minimum and its second peak are shifted to larger positions as compared to those for the RIM (and $g_{++}$ is not in phase with $g_{--}$); and (b) the first peak of $g_{+-}$ for PIM1s (whose value is 3.32) is lower than that for RIM (3.64), and it is shifted from 2.5 Å to 2.75 Å. The first peak of $g_{+-}$ for PIM1, at 2.6 Å, is lower (2.90) than those for RIM and PIM1s. These small differences, besides the induced polarization effects, are also due to the differences between $f_{ab}^{o}(r)$ and $f_{ab}^{VR}(r)$.

We now turn to the static structure factors. The Ashcroft–Langreth partial structure factors $S_{ab}(k)$ are shown in Figure 3. As it is well known, the charge ordering is reflected in reciprocal space by the principal peaks and the valley of $S_{ab}(k)$ at the same wave-number (about 1.8 Å$^{-1}$). As expected from the $g_{ab}(r)$ results discussed above, the most salient differences in comparing the three models are between the $S_{++}$ for PIM1 and those for RIM and PIM1s. The principal peak of $S_{++}$ for PIM1 is lower than those for RIM and PIM1s. Moreover, the $S_{++}$ for PIM1 shows a second peak at about 2.3 Å$^{-1}$, which is decisive in accounting for the three-peak feature in the broad main peak of the total structure factor (see below). This second peak is related to the *wavelength* $l = 2.75$ Å between the second and third peaks of $g_{++}$, at 5.5 Å and 8.25 Å respectively. Concerning the PIM1s it is also worth noting the shoulder of $S_{++}$ at about 2.2 Å$^{-1}$. This shoulder indicates again that the short-range polarization interactions included in PIM1s partially cancel the effects of the electrically induced dipoles. In addition, the $S_{--}$ for the three models are very similar, while the principal valley of $S_{+-}$ for PIM1 is shallower (−1.38) than that for RIM (−1.64) and PIM1s (−1.96). Then, the first sharp peak of $S_{ZZ}=[S_{++}+S_{--}-2S_{+-}]/2$ is lower for PIM1 (3.59) than that for RIM (4.16) and PIM1s (4.59), reflecting that charge ordering is less marked in the PIM1. The Bhatia–Thornton structure factors are plotted in Figure 4.

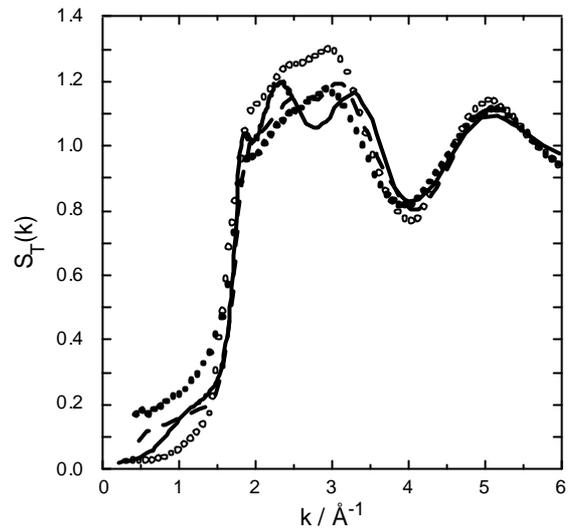

**Figure 5.** Neutron scattering data for the total static structure factor, $S_T(k)$, of molten AgBr obtained by Inui et al.[5] at 753 K (open circles), Keen et al.[27] at 703 K (dashed line) and *Takeda* et al.[28] at 773 K (full circles). MD results of this work for the $S_T(k)$ of molten AgBr at 753 K using the PIM1 (solid line).

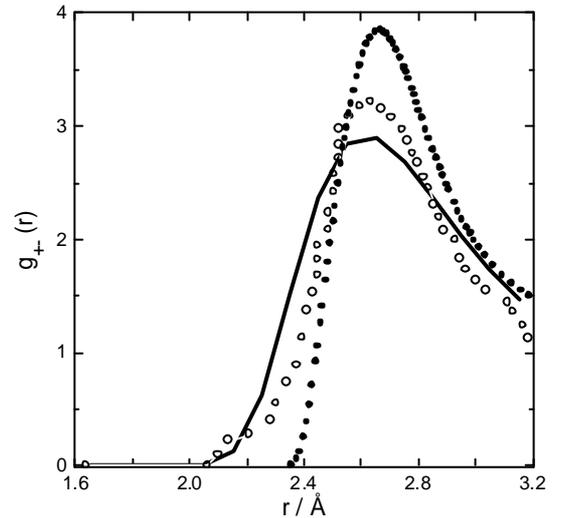

**Figure 6.** First peak of $g_{+-}(r)$ for molten AgBr determined by Keen et al.[27] from reverse Monte Carlo data analysis of neutron diffraction experiments at 703 K (open circles); by Di Cicco et al.[35] after data analysis of EXAFS measurements at 725 K (full circles); and from MD simulation at 753 K using the PIM1 (solid line), this work.

As in the molten alkali halides, $S_{NN}=[S_{++}+S_{--}+2S_{+-}]/2$ is less structured than $S_{ZZ}$, with a rather low and broad peak between the first and second peaks of $S_{ZZ}$. However, while in the molten alkali halides the broad main peak of $S_{NN}$ is at about the same wave-number as the first maximum of $S_{+-}$,[33] for the RIM and the PIM1s it is decoupled into two relative maxima. On the other hand, $S_{NN}$ for PIM1 is decoupled into a shoulder, at 1.9 Å$^{-1}$, and two maxima. The first maximum for the RIM, at about the same wave-number as the principal peak of $S_{ZZ}$, is less pronounced than the second, at lower $k$ values than the second peak of $S_{ZZ}$. For the PIM1s it is exactly the opposite with the first maximum at higher $k$ values. The shoulder and the second maximum of $S_{NN}$ for PIM1 are at approximately the same $k$ values as the RIM first and second maxima, while the middle maximum is at 2.35 Å$^{-1}$. The shape of $S_{NN}$ is due to subtle cancellations between the features present in the partial structure factors, and reflects the differences between the partials discussed above. Furthermore, since the values of neutron scattering lengths for normal isotopic composition of the Ag and Br ions are relatively similar ($b_+ = b_{Ag}^{nat} = 5.922$ fm and $b_- = b_{Br}^{nat} = 6.795$ fm),[34] the total structure factor $S_T = [b_+^2 S_{++} + b_-^2 S_{--} + 2b_+ b_- S_{+-}]/(b_+^2 + b_-^2)$ is very similar to $S_{NN}$. Hence, $S_T$ reflects the topological order present in $S_{NN}$. A careful comparison between $S_{NN}$ and $S_T$ for PIM1 shows that the shoulder of $S_{NN}$ at 1.9 Å$^{-1}$ becomes a relative maximum in $S_T$. Therefore, the broad main peak of $S_T$ presents three maxima in the PIM1. In Figure 3 we show the $S_T$ together with the partials. Here, in the middle panel, we see that the second relative maximum of $S_T$ for the PIM1, at 2.35 Å$^{-1}$, is almost at the same wave-number as the second peak of $S_{++}$, at 2.30 Å$^{-1}$.

**TABLE 3: MD Results of the Self-Diffusion Coefficients and the Conductivity for Molten AgBr at 753 K**[a]

|  | RIM | PIM1 | PIM1s |
|---|---|---|---|
| $D_+ / 10^{-5}$ cm$^2$/s | 3.85 | 5.00 | 2.34 |
| $D_- / 10^{-5}$ cm$^2$/s | 2.00 | 1.42 | 1.09 |
| $\sigma / (\Omega\cdot\text{cm})^{-1}$ | 3.3 | 2.8 | 1.9 |
| $\sigma_{NE} / (\Omega\cdot\text{cm})^{-1}$ | 2.56 | 2.8 | 1.5 |
| $\Delta$ | −0.30 | ≈ 0 | −0.28 |

[a] Experimental value[26] $\sigma$ = 2.97 $(\Omega\cdot\text{cm})^{-1}$.

From the simulated models, the $S_T$ that better compares to the available neutron scattering data for molten AgBr is that of the PIM1, even though the smoothed three-peak feature in the broad main peak is overestimated. The comparison between the calculated $S_T$ for PIM1 and the experimental data is shown in Figure 5. The three total scattering data are those measured by Inui et al.[5] at 753 K, Keen et al.[27] at 703 K, and Takeda et al.[28] at 773 K. The broad main peak of the first is higher and broader than the other two, with the latter the lower and narrower. Nevertheless, within the broad peak, following the first feature at about 1.9 Å$^{-1}$, the three experimental $S_T$ increase slowly, with a mid shoulder, until the highest peak. Since these features resemble the characteristic three-peak structure of molten AgCl, we refer to them as the smoothed three-peak feature of the broad mean peak.

Keen et al.[27] analyzed their $S_T(k)$ using the Reverse Monte Carlo (RMC) modeling technique to produce the three radial distribution functions $g_{ab}(r)$. Furthermore, Di Cicco et al.[35] determined the first peak of $g_{+-}$ for molten AgBr at 725 K after a sophisticated data analysis of their EXAFS measurements. In Figure 6 we compare both the RMC and EXAFS data analysis results for the first peak of $g_{+-}$ with our MD results using the PIM1. The highest and sharpest peak of $g_{+-}$ corresponds to the EXAFS data analysis. Moreover, this peak is the less symmetric with a sharp drop at the left side. The position of the first peak for PIM1, at 2.6 Å, is close to those from the RMC, at 2.63 Å, and EXAFS, at 2.67 Å, but its height (2.90) is lower. The first peak of $g_{+-}$ for RIM (3.64) is higher than that for PIM1s (3.32), but the former is shifted to lower distances, at 2.5 Å, while the latter is at 2.75 Å.

**4.2. Ionic Transport Properties** The relevant time correlation functions to describe the averaged single-ion motion are the self-velocity autocorrelation functions,

$$\Lambda_a(t) = \frac{1}{3}\langle \mathbf{v}_{i_a}(t)\cdot\mathbf{v}_{i_a}(0)\rangle = \frac{1}{3}\frac{1}{N_a}\sum_{i_a=1}^{N_a}\langle \mathbf{v}_{i_a}(t)\cdot\mathbf{v}_{i_a}(0)\rangle \quad (4.3)$$

where $\mathbf{v}_{i_a}$ is the velocity of particle $i_a$ of species $a$, and the mean square displacements of the single ions

$$Q_a(t) = \frac{1}{3}\langle |\mathbf{r}_{i_a}(t) - \mathbf{r}_{i_a}(0)|^2\rangle = \frac{1}{3}\frac{1}{N_a}\sum_{i_a=1}^{N_a}\langle |\mathbf{r}_{i_a}(t) - \mathbf{r}_{i_a}(0)|^2\rangle \quad (4.4)$$

from which the self diffusion coefficients $D_a$ can be evaluated using both the Kubo and Einstein formula

$$D_a = \int_0^\infty \Lambda_a(t)dt \quad \text{and} \quad D_a = \lim_{t\to\infty} \frac{Q_a(t)}{2t} \quad (4.5)$$

The basic collective time correlation functions to study the ionic conduction in binary melts are the charge-current density autocorrelation function[36,12]

$$\Lambda_{ZZ}(t) = \frac{1}{3}N\langle \mathbf{j}_Z(t)\cdot\mathbf{j}_Z(0)\rangle \quad \text{with} \quad \mathbf{j}_Z = \frac{1}{N}\sum_{i=1}^N z_i\mathbf{v}_i \quad (4.6)$$

and the mean square displacement of the charge-center position[37,38]

$$Q_{ZZ}(t) = \frac{1}{3}N\langle |\mathbf{r}_Z(t) - \mathbf{r}_Z(0)|^2\rangle \quad \text{with} \quad \mathbf{r}_Z = \frac{1}{N}\sum_{i=1}^N z_i\mathbf{r}_i \quad (4.7)$$

from which the ionic conductivity $\sigma$ can be evaluated by the corresponding Kubo and Einstein-like relations

$$\sigma = \frac{r_N e^2}{k_B T}\int_0^\infty \Lambda_{ZZ}(t)\,dt \quad \text{and} \quad \sigma = \frac{r_N e^2}{k_B T}\lim_{t\to\infty}\frac{Q_{ZZ}(t)}{2t} \quad (4.8)$$

where $k_B$ is the Boltzmann constant. While $D_a$ are single particle (self) properties, $\sigma$ is a collective property that requires to be averaged over long simulations. The length of the simulations in this work is of $3\times 10^5$ time steps, and the estimated uncertainties for the values of $\sigma$ are about 10%, whereas they are less than 5% for $D_a$. The accordance between the values calculated by using the Kubo and the Einstein relations are within the error intervals.

In Table 3 we show the results for the self-diffusion coefficients $D_a$, and the conductivity $\sigma$, of molten AgBr at 753 K. In all cases the cations are more mobile than the anions –with $D_+$ about two, or more, times larger than $D_-$– showing the superionic character of the melt. Moreover, we see that the cation PIM1 diffusion coefficient is larger, and the anion smaller, than the corresponding RIM values; with the PIM1s results smaller in both cases. The largest cation diffusivity for PIM1 is consistent with the $g_{++}$ results; since the $g_{++}$ for PIM1 is the less structured and presents the deepest cation penetration. It is as if the overall effect of the electrically induced dipoles in the anions increases the free space for the cations that can diffuse through "faster" channels. However, this effect is canceled by the short-range damping interactions in the PIM1s and, therefore, the corresponding $D_+$ is smaller. Moreover, since the repulsive wall of the pair potential $f^0_{+-}(r)$ is at longer distances than that of the pair potential $f^{VR}_{+-}(r)$ assumed in the RIM, the free space for the cations is smaller in the PIM1s and the corresponding $D_+$ is lower than that for RIM. It is worth noting that the $D_+$ values are larger as the charge ordering reflected by the first sharp peak of $S_{ZZ}$ is less marked. However, this simple rule is not satisfied by the $D_-$ values. We suspect this may be due to the differences between $f^0_{ab}(r)$ and $f^{VR}_{ab}(r)$.

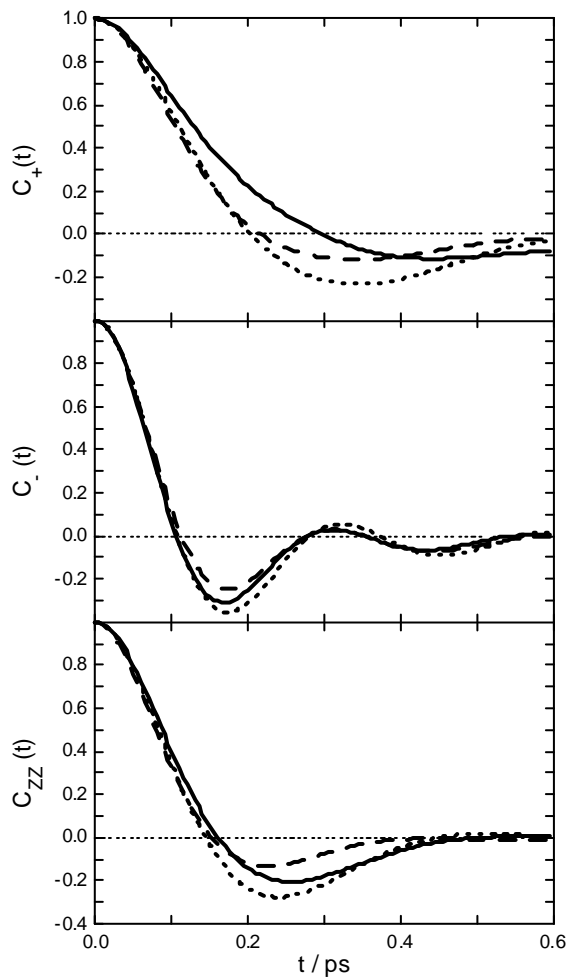

**Figure 7.** Normalized self-velocity autocorrelation functions, $C_a(t)=\Lambda_a(t)/\Lambda_a(0)$, for cations (top) and anions (middle), as well as $C_{ZZ}(t)=\Lambda_{ZZ}(t)/\Lambda_{ZZ}(0)$ (bottom), from MD simulations using the following models: RIM (dashed line), PIM1 (solid line) and PIM1s (dotted line).

The normalized self-velocity autocorrelation functions, $C_a(t)=\Lambda_a(t)/\Lambda_a(0)$, are compared in Figure 7. In all cases the $C_-(t)$ are oscillatory after the initial decay and the $C_+(t)$ have a slower decay and show a weaker backscattering. This behavior is because the cations are heavier and smaller than the anions.[39] Concerning the cations averaged motion, the $C_+(t)$ for PIM1 shows the slowest decay and the weakest backscattering, whereas the minima of the $C_+(t)$ for RIM and PIM1s are at about the same time, with the latter the deeper. On the other hand, the three $C_-(t)$ oscillate at almost the same frequency, with the most pronounced oscillations in the PIM1s. At this point we recall the basic mechanism for the averaged ionic motion suggested in previous papers for the copper and silver halides melts.[39]

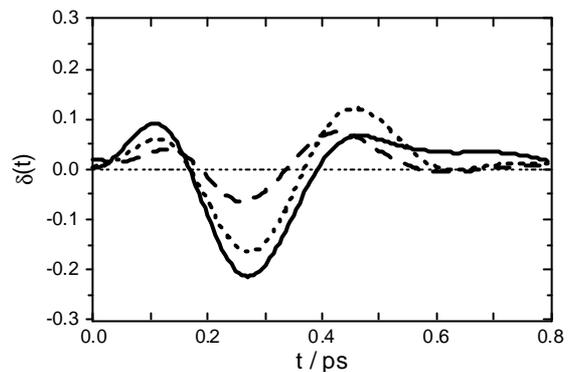

**Figure 8.** Distinct time correlation function $d(t)$ from MD simulations of molten AgBr at 753 K using the following models: RIM (dashed line), PIM1 (solid line) and PIM1s (dotted line).

While the anions, much larger than the cations, experience a rattling motion in a double cage formed by neighboring unlike and like ions, the cations diffuse through the packed structure of the slowly diffusing anions. Thus, the more pronounced solid-like oscillatory behavior of the anions in the PIM1s and the PIM1 suggests that their effective size is larger than that in the RIM, while the more fluid-like diffusive behavior of the cations in the PIM1s suggests that their effective size is smaller than that in the RIM and the PIM1s. All these features are consistent with the self-diffusion results discussed above.

As in previous papers,[1,12] we carried out the conductivity calculations using the formal charge $|z_a|=1$. Whereas in their interactions we assume the ions only "see" effective charges, which we tentatively attribute to an average screening effect of the electronic shells, in their transport we assume the ions carry with them the full complement of electrons. The value of the conductivity obtained for the PIM1 (2.8 $\Omega^{-1}$cm$^{-1}$) is in good agreement with the experimental values (2.97 $\Omega^{-1}$cm$^{-1}$)[26], while it is larger for RIM (3.3 $\Omega^{-1}$cm$^{-1}$) and smaller for PIM1s (1.9 $\Omega^{-1}$cm$^{-1}$). The normalized charge-current autocorrelation functions $C_{ZZ}(t)=\Lambda_{ZZ}(t)/\Lambda_{ZZ}(0)$, which are compared in Figure 7, are consistent with the $s$ values. The deeper the minimum, the larger the negative area under $C_{ZZ}(t)$ and lower the conductivity.

The conductivities are often estimated from the self-diffusion coefficients by the Nernst–Einstein approximation which, for binary melts with $z_+=z_-=1$, is written[33,36]

$$\boldsymbol{s}_{NE} = \frac{r_N e^2}{k_B T}\frac{1}{2}(D_+ + D_-) \quad \text{and} \quad \boldsymbol{s} = \boldsymbol{s}_{NE}(1-\Delta) \qquad (4.9)$$

where $-\Delta \boldsymbol{s}_{NE}$ is the contribution to $\boldsymbol{s}$ of the cross-correlations between the velocities (or displacements) of different ions. These correlations are given by the distinct time correlation function

$$\boldsymbol{d}(t) = \frac{\Lambda_{ZZ}(t) - \frac{1}{2}[\Lambda_+(t)+\Lambda_-(t)]}{\Lambda_{ZZ}(0)} \qquad (4.10)$$

which is normalized by $\Lambda_{ZZ}(0)$ because $\Lambda^d(0)/4=\Lambda_{ZZ}(0)-[\Lambda_+(0)+\Lambda_-(0)]/2=0$,[40] and

$$\Delta = -\frac{2\Lambda_{ZZ}(0)}{D_+ + D_-}\int_0^\infty \boldsymbol{d}(t)dt \qquad (4.11)$$

In Figure 8 we plotted $\boldsymbol{d}(t)$, and the values of $\Delta$ are shown in

Table 3. Since $d(t)$ starts at zero and oscillates around 0, and $\Delta$ is proportional to the area under $d(t)$, $\Delta$ depends on subtle cancellations between the positive and negative contributions of $d(t)$.

Looking at Table 3, we note that, relative to $s$, the $s_{NE}$ values are lower for RIM and PIM1s ($\Delta<0$), and practically the same for PIM1 ($\Delta\approx0$). The present results for AgBr using the PIM1, as well as those for molten AgCl (with $\Delta>0$),[12] reverse the systematic negative values of $\Delta$ obtained in our previous calculations for copper and silver halide melts using the RIM with Vashishta and Rahman pair potentials.[41,1] Unfortunately, to our knowledge, with the exception of molten CuCl,[42] there are no experimental data on the self-diffusion coefficients for these superionic melts, and the MD results of $\Delta$ can not be compared to experimental values. We recall that in molten alkali halides it is found that $\Delta>0$.[43] Nevertheless, we note in Figure 8 that in all cases $d(t)$ starts from zero to positive values. This behavior has also been found in previous MD simulations of copper halide melts using rigid ion models as well as for AgCl using both rigid and polarizable ion models.[44,12] Since this behavior differs from that found for alkali halide melts, where $d(t)$ always starts from zero to negative values,[44] it can be considered as another benchmark of the superionic character of the melt.

**5. Conclusions**

In this work we have studied different polarizable ion models of molten AgBr based on a rigid ion model (RIM) with the pair potential of the type proposed by Vashishta and Rahman, and assuming that the dipoles can be induced by (a) the local electric field due to all other ions and (b) the short-range overlap effects due to the neighboring ions. We have denoted the models with the ions polarizable solely by the local electric field as PIM1 and PIM2. In the former only the anions are polarizable, whereas in the latter both cations and anions are polarizable. The models that include the short-range polarization effects have been denoted as PIM1s and PIM2s.

It is worth mentioning that the polarizable ion models based on the pair potential of the Born–Mayer form proposed for AgBr by Mayer[10] fail to predict the experimental melting point, and the corresponding systems do not melt at the thermodynamic state at which experimental neutron scattering data are available.

It has been shown that the PIM2 is an unphysical model since it leads to the polarization catastrophe, i.e. the ions become infinitely polarized when they approach a critical distance. Moreover, we have found that the short-range damping polarization interactions used in this work do not guarantee the stability of the PIM2s, since there are some improbable, but possible, critical configurations where the polarization catastrophe also occurs. We believe that these critical configurations could be avoided by using short-range damping interactions that also oppose the electric field due to the neighboring induced dipoles in the same way as, for instance, the damping procedures proposed for water models by Stillinger and Davids,[45] Halley et al.,[46] or Thole.[24]

To study the influence of both the electrical and short-range damping polarization effects on the properties of molten AgBr, we have carried out molecular dynamics simulations using the PIM1, the PIM1s, as well as the RIM.

From the results we find that, although we only take into account the anion polarizability, it is mainly the cations that are affected by the induced polarization interactions. This may be attributed to the screening of the cations repulsion due to the anion induced dipoles. Moreover, we find that the short-range damping interactions included in PIM1s partially cancel, as expected, the effects of the electrically induced dipoles.

The most relevant results are those obtained from PIM1. The PIM1 cations are less structured and more diffusive than those of RIM and PIM1s, while the anion-anion structure shown by $g_{--}$ and $S_{--}$ is practically the same for the three models. It is as if the overall effect of the induced dipoles in the anions increases the free space for the cations which can diffuse through "faster" channels. However, this effect is canceled by the short-range polarization effects in PIM1s, and the corresponding $D_+$ is lower than that from RIM. The PIM1 reproduces the broad main peak of the total structure factor present in the available neutron diffraction data, although the smoothed three-peak feature of this broad peak is overestimated, while the RIM and the PIM1s fail to reproduce these features and decouple the broad peak into two peaks. Moreover, the PIM1 conductivity is in good agreement with experimental values.

Although the better agreement with experimental data is obtained using the PIM1, we believe that short-range polarization effects cannot be neglected. Probably, PIM1s simulations with a lower damping parameter $r_{ab}$ will also reproduce the characteristic smoothed three-peak structure of molten AgBr, as well as improve the conductivity, but we have not checked this point yet. Moreover, we note that the values of the ionic transport coefficients, besides the induced polarization effects, are strongly determined by the pair potential $f_{ab}^0(r)$. Thus, small corrections in the parameterization of $f_{ab}^0(r)$ that retain the most salient features of the structure, could change the conductivity in the range of the values obtained using the three different models.

**Acknowledgements**. It is a pleasure to thank M. Silbert for fruitful discussions regarding this work. This work was supported by DGICYT of Spain (BFM2003-08211-C03-01), the DURSI of the Generalitat of Catalonia (2001SGR-00222), and European Union FEDER funds (UNPC-E015). One of us (VB) also thanks the Spanish Ministry of Education and Science for FPU grant AP2003-3408.